\documentclass[pra,reprint,superscriptaddress]{revtex4-1}

\pdfoutput=1

\usepackage{amsmath,amssymb, graphicx, natbib, xcolor}
\usepackage{stackengine}[2013-09-11]
\usepackage{amsthm}
\newtheorem{theorem}{Theorem}
\newtheorem{definition}{Definition}[section]

\newcommand{\+}{^\dagger}
\newcommand{\nodag}{^{\phantom\dagger}}
\newcommand{\hH}{\hat{H}}
\newcommand{\hb}{\hat{b}}
\newcommand{\hc}{\hat{c}}
\newcommand{\ket}[1]{\left| #1 \right\rangle}
\newcommand{\bra}[1]{\left\langle #1 \right|}
\newcommand\slashzero{\emptyset}

\begin{document}
\author{Bhuvanesh Sundar}
\email{Bhuvanesh.Sundar@rice.edu}
\affiliation{Department of Physics and Astronomy, Rice University, Houston, Texas 77005, USA}
\affiliation{Rice Center for Quantum Materials, Rice University, Houston, Texas 77005, USA}

\author{Matthew Thibodeau}
\email{Matt.Thibodeau@rice.edu}
\affiliation{Department of Physics and Astronomy, Rice University, Houston, Texas 77005, USA}

\author{Zhiyuan Wang}
\email{Zhiyuan.Wang@rice.edu}
\affiliation{Department of Physics and Astronomy, Rice University, Houston, Texas 77005, USA}
\affiliation{Rice Center for Quantum Materials, Rice University, Houston, Texas 77005, USA}

\author{Bryce Gadway}
\email{bgadway@illinois.edu}
\affiliation{Department of Physics, University of Illinois at Urbana Champaign, Urbana, Illinois 61801, USA}

\author{Kaden R. A. Hazzard}
\email{kaden@rice.edu}
\affiliation{Department of Physics and Astronomy, Rice University, Houston, Texas 77005, USA}
\affiliation{Rice Center for Quantum Materials, Rice University, Houston, Texas 77005, USA}

\title{Strings of ultracold molecules in a synthetic dimension}
\date{\today}

\begin{abstract}
We consider ultracold polar molecules trapped in a unit-filled one-dimensional chain in real space created with an optical lattice or a tweezer array and illuminated by microwaves that resonantly drive transitions within a chain of rotational states. We describe the system by a two-dimensional lattice model, with the first dimension being a lattice in real space and the second dimension being a lattice in a synthetic direction composed of rotational states. We calculate this system's ground-state phase diagram. We show that as the dipole interaction strength is increased, the molecules undergo a phase transition from a two-dimensional gas to a phase in which the molecules bind together and form a string that resembles a one-dimensional object living in the two-dimensional (i.e., one real and one synthetic dimensional) space. We demonstrate this with two complementary techniques: numerical calculations using matrix product state techniques and an analytic solution in the limit of infinitely strong dipole interaction. Our calculations reveal that the string phase at infinite interaction is effectively described by emergent particles living on the string and that this leads to a rich spectrum with excitations missed in earlier mean-field treatments.
\end{abstract}

\maketitle
\section{Introduction}\label{sec: intro}
Ultracold polar molecules have been predicted to show strongly correlated phases~\cite{wall2015quantum, gorshkov2011tunable, gorshkov2011quantum, gorshkov2013kitaev, manmana2017correlations, manmana2013topological, lemeshko2013manipulation, moses2017new, gadway2016strongly, carr2009cold, fedorov2016novel, brennen2007designing, hazzard2013far, micheli2006toolbox, barnett2006quantum, wall2013simulating, wall2015realizing} arising from the rich structure of their internal quantum states~\cite{koch2018quantum, krems2018molecules} and strong dipole interactions between them. Motivated by this, researchers have created degenerate~\cite{de2019degenerate} or nearly degenerate gases~\cite{ni2008high, takekoshi2014ultracold, molony2014creation, park2015ultracold, guo2016creation, seesselberg2018extending} of various ground-state polar molecules such as KRb, NaRb, NaK, and RbCs. They have observed the effect of the dipole interactions on the dynamics of the molecules' rotational states~\cite{hazzard2014many, yan2013observation}.

A recent paper~\cite{sundar2018synthetic} by some of the present authors proposed that one can utilize the rotational states of polar molecules to realize a fully controllable synthetic lattice with a large number of sites. The synthetic lattice sites are the molecule's rotational states, and synthetic tunnelings are driven by resonant microwaves. The synthetic lattice in this setting can potentially have over a hundred sites, and the tunneling elements and the energy landscape can be fully controlled via the microwave fields. It was argued that realizing a synthetic lattice this way can have potential advantages over other experimental realizations (such as Refs.~\cite{an2017ballistic, an2017correlated, mancini2015observation, stuhl2015visualizing, celi2014synthetic, anisimovas2016semisynthetic, wall2016synthetic, livi2016synthetic, kolkowitz2017spin, floss2013anderson, floss2015observation}), most notably in the synthetic lattice size that can be realized, as well as potentially lower susceptibility to magnetic field noise.

Further, Ref.~\cite{sundar2018synthetic} argued that molecules trapped in a lattice along one synthetic and $d=1$ or 2 real directions exhibit strongly correlated many-body phases. Specifically, they argued using mean-field theory that sufficiently strong dipole interactions between molecules cause the molecules to undergo a phase transition from a $(d+1)$-dimensional gaslike phase into a phase where the molecules bind into a $d$-dimensional object, remarkably undergoing a collapse in the synthetic dimension. For $d=1$, the bound phase is an effective one-dimensional (1D) object fluctuating in a two-dimensional (2D) space, and thus is a ``quantum string.'' Similarly for $d=2$, the bound phase is a ``quantum membrane.''

However, the mean-field theory in Ref.~\cite{sundar2018synthetic} left many questions unanswered. Foremost among them is the validity of the mean-field theory. This question is especially acute for $d=1$ with strong interactions, given the known limitations of mean-field theory. Reference~\cite{sundar2018synthetic} also left open the nature of the excitations of the ground state.

In this paper, we calculate the ground state of ultracold polar molecules in one real and one synthetic dimension, first numerically using matrix product state (MPS) methods~\cite{dukelsky1998equivalence, jaschke2017open} and then analytically in the limit of strong dipole interactions. We show numerical evidence that the system has two string phases and an unbound phase, consistent with the mean-field theory of Ref.~\cite{sundar2018synthetic}. We show that in the limit of infinitely strong dipole interactions, the ground states and excitation spectrum map to those of a hardcore boson model. We analytically calculate the ground-state wave function in this limit and show that the system spontaneously forms strings that are two or three synthetic sites wide.

While our results confirm the basic mean-field picture of two string phases and a gaslike phase, they also reveal additional effects missed by the mean-field theory. The most notable is the existence of exponentially many width-3 strings in the limit of infinitely strong dipole interactions. The mean-field theory predicts only the width-2 strings. Our results also imply an intricate excitation spectrum, including the fluctuations within width-2 strings as well as between the width-2 and width-3 strings, all of which were absent in the mean-field theory. Though these are zero-energy excitations due to the degeneracy in the infinitely strong interaction limit, they presumably become gapless excitation modes at finite dipolar interaction strength. There could be a concern that these numerous low-energy excitations may make the ground state sensitive to perturbations. To explore this, we numerically calculate the ground state in the presence of modified interactions which are predicted to gap some of the gapless excitations, and show that the string ground states indeed constitute a stable phase.

This article is organized as follows. In Sec.~\ref{sec: model}, we describe the setup. In Sec.~\ref{sec: dmrg}, we numerically explore the phase diagram using MPS methods, defining and calculating a correlation that characterizes the string and gas phases. In Sec.~\ref{sec: exact}, we analytically derive the ground-state wave function in the limit of infinitely strong dipole interactions, show that they have a finite synthetic width, and expose the richness of the excitation spectrum at this point. In Sec.~\ref{sec: electric}, we calculate the ground states in the presence of modified interactions, and show that the ground-state strings are robust and seem to be stabilized by certain kinds of interactions. In Sec.~\ref{sec: summary}, we summarize our findings.
\begin{figure}[t]\centering
\begin{tabular}{c}
\begin{tabular}{ccc}
\includegraphics[width=0.55\columnwidth]{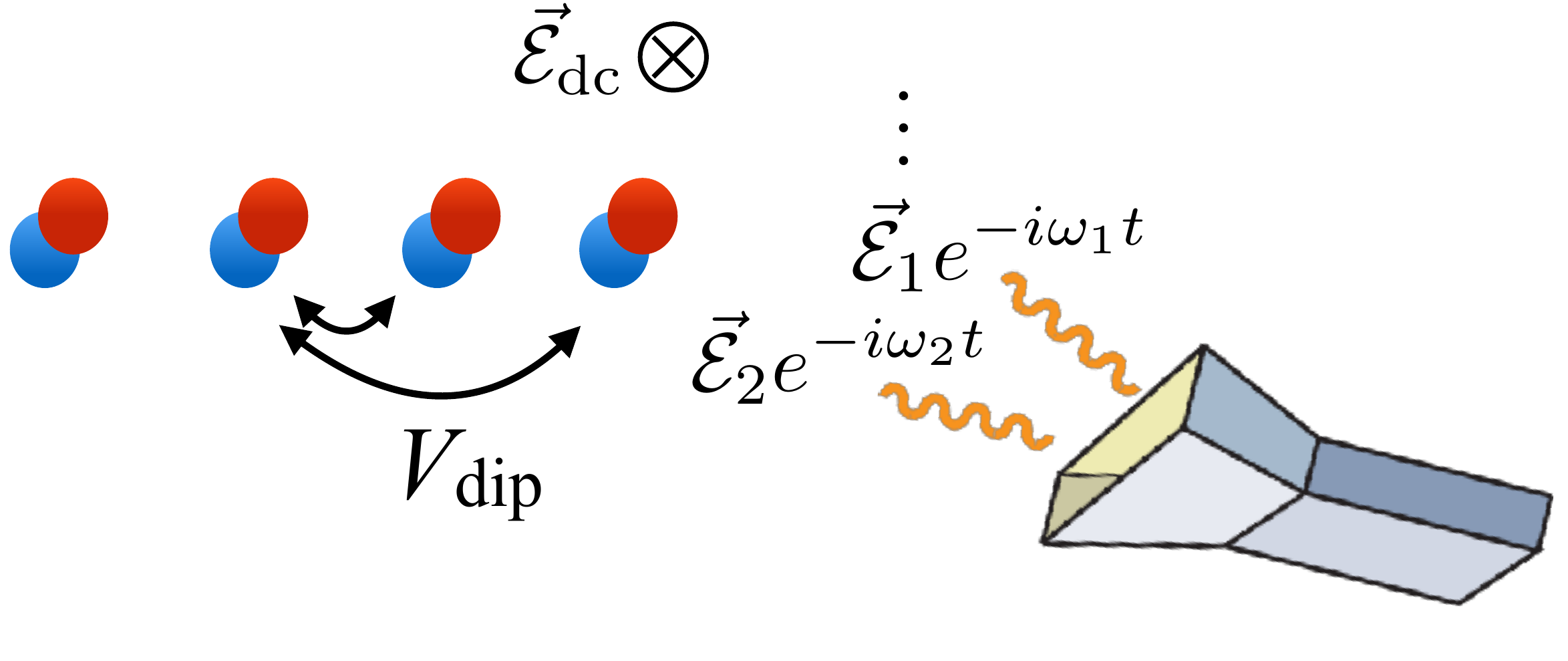} && \includegraphics[width=0.35\columnwidth]{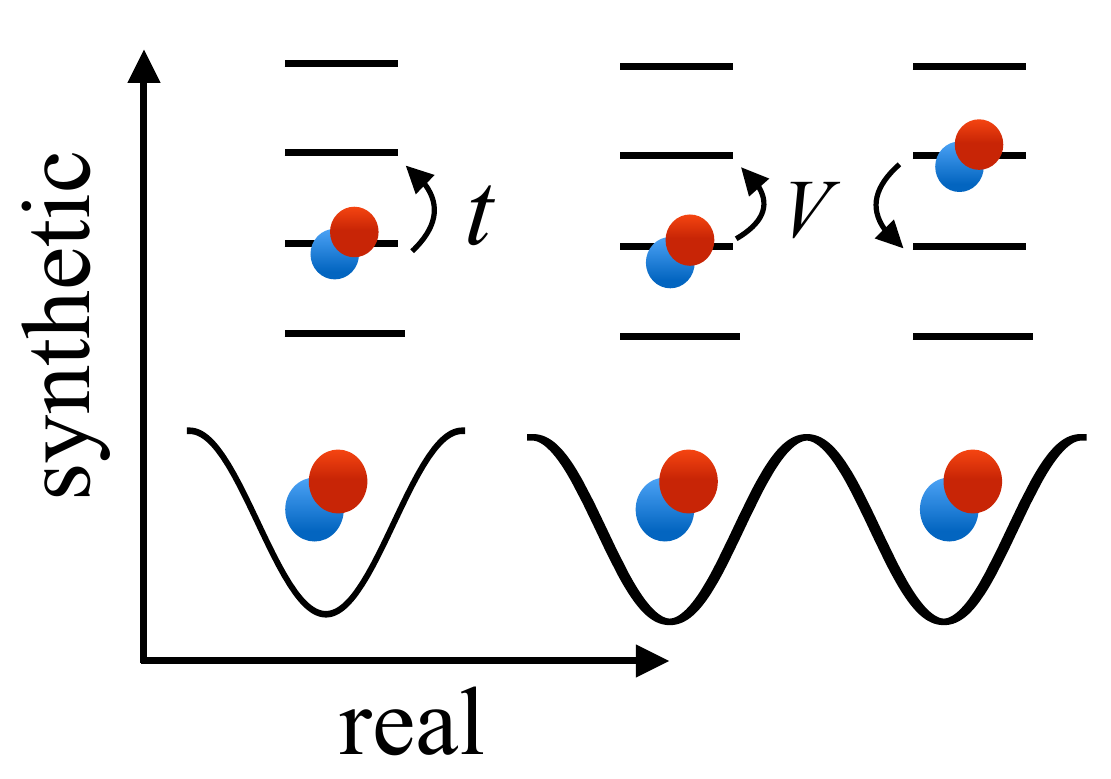}\\
(a) && (b)
\end{tabular}\\ \\
\includegraphics[width=0.95\columnwidth]{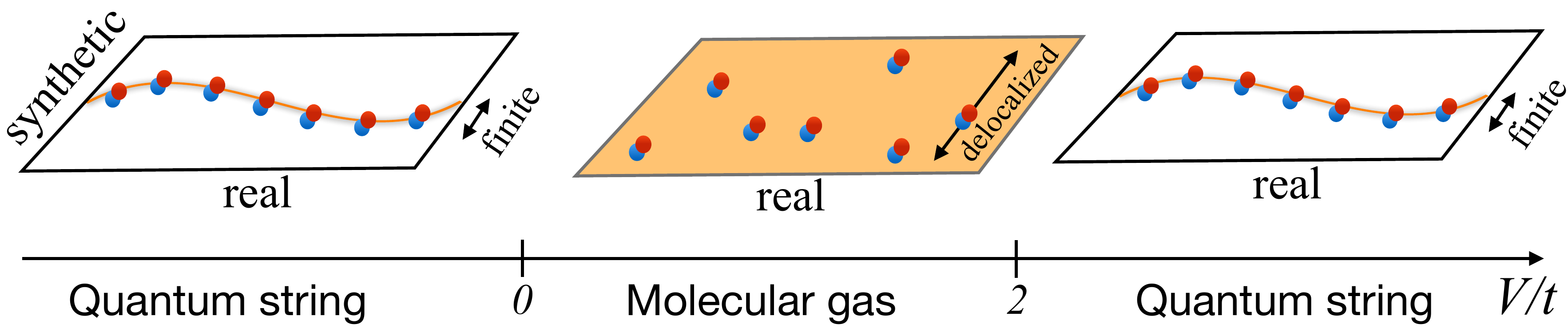}
\\(c)
\end{tabular}
\caption{(Color online) (a) Schematic illustration of the system, consisting of a 1D real-space unit-filled periodic array of polar molecules, with several microwaves illuminating them. A small dc electric (or magnetic) field maybe required to detune away undesired molecular transitions. (b) The physical processes in the system. The synthetic tunneling $t>0$ is driven by microwaves. The dipole interaction between molecules takes the form of a correlated tunneling process with strength $V$. The origin of all these processes is explained in Ref.~\cite{sundar2018synthetic}. (c) The main result of this work: The many-body ground state is a ``quantum string'' bound state of molecules for $V/t
\lesssim0$ and $V/t\gtrsim2$, and an unbound molecular gas otherwise.}
\label{fig: setup}
\end{figure}

\section{Setup}\label{sec: model}
We consider a unit-filled periodic array of molecules trapped in a 1D real space chain, created using a deep optical lattice or a tweezer array~\cite{kaufman2012cooling, schlosser2012fast, piotrowicz2013two, nogrette2014single, lester2015rapid, barredo2016atom, endres2016atom, weitenberg2011quantum, labuhn2016tunable, kim2016situ,  bernien2017probing, levine2018high, norcia2018microscopic, cooper2018alkaline}. This setup, first introduced in Ref.~\cite{sundar2018synthetic}, is illustrated in Fig.~\ref{fig: setup}(a). We consider the molecules to be in their lowest electronic and vibrational state and a single hyperfine state, and denote their rotational states as $\ket{n,m}$ with rotational angular momentum $n\hbar$ and azimuthal angular momentum $m\hbar$ about the $z$ axis. The molecules are illuminated with microwaves that resonantly drive transitions between a subset of the rotational states in the full rotational spectrum. This subset can be visualized as a synthetic lattice, and the rotational state transitions as synthetic tunnelings. The specific rotational states that make the synthetic lattice can be controlled by tuning the microwave polarizations and frequencies appropriately, and adding a small dc electric or magnetic field to detune away transitions to other rotational states. For example, a 1D synthetic lattice with lattice sites as the rotational states $\ket{n,0}$ can be created by shining $\Pi$-polarized microwaves and applying an electric field along $\hat{z}$. In this paper, we consider a 1D synthetic lattice and label the synthetic sites (say, $\ket{n,0}$) by the index $n$ running up to $N_{\rm int}-1$, with $N_{\rm int}$ being the synthetic lattice size. This is an extension of the setup considered in Refs.~\cite{gorshkov2011tunable, hazzard2013far, yan2013observation, hazzard2014many} with $N_{\rm int}=2$; however, in those works, it was more convenient to describe the molecular states as spin-1/2 instead of a synthetic lattice with two sites.

The Hamiltonian for the molecular array, in the frame rotating at the microwaves' frequencies, is
\begin{align}
\hH =  & -t\sum_{j=0}^{N_{\rm real}-1}\sum_{n=1}^{N_{\rm int}-1} \hc_{n-1,j}\+\hc_{nj}\nodag + V\sum_{n \langle ij\rangle} \hc_{n-1,i}\+\hc_{ni}\nodag\hc_{nj}\+\hc_{n-1,j}\nodag \nonumber\\ &+ {\rm H.c.},
\label{eqn: H}
\end{align}
where $\hc_{nj}\nodag$ ($\hc_{nj}\+$) annihilates (creates) a molecule on real site $j$ and synthetic site $n$. $N_{\rm real}$ and $N_{\rm int}$ are the real and synthetic lattice sizes. We emphasize that $N_{\rm int}$ can potentially be a hundred or even larger, making polar molecules attractive candidates to realize a synthetic lattice.

Figure~\ref{fig: setup}(b) illustrates the physical processes in the system. The synthetic tunneling $t$ is driven by microwaves. We consider the microwaves to be tuned such that the synthetic tunnelings are uniform and choose a gauge such that $t>0$. Dipole interactions cause a pair of molecules to resonantly swap between adjacent synthetic sites, with a strength $V$. The sign of $V$ can be adjusted via the orientation of the molecular chain relative to the quantization axis, or appropriately choosing the synthetic lattice sites in the rotational spectrum. For example, if the synthetic lattice sites are the $\ket{n,0}$ states, then $V>0$ if the molecular chain is on the $x$-$y$ plane, and $V<0$ if the chain is along $\hat{z}$. We assume that the real space lattice is sufficiently deep that real space tunnelings are negligible, and together with unit filling, this imposes the constraint $\sum_n \hc_{nj}\+\hc_{nj}\nodag=1\ \forall j$.

We neglect the effects of the long-range nature of the dipole interactions, considering only nearest-neighbor interactions in real space. We do not expect this to significantly alter the physics in a system with one real dimension. The strength of the swap process depends on $n$, but this dependence is weak and nearly independent of $n$ for large $n$, so it can be eliminated by working with sufficiently excited rotational states. We have explicitly checked within mean-field theory that neither the long-range dipolar interactions nor the dependence of $V$ on $n$ affect the physics substantially. Instead, they only shift the phase boundaries.

\subsection{Special cases}\label{subsec: spl cases}
The ground states of $\hH$ can be calculated analytically at least for three cases: (i) $V=0$, (ii) $N_{\rm real}=2$ and $N_{\rm int}\rightarrow\infty$, and (iii) $t=0$. We briefly review these cases here for context.

The solution for (i), $V=0$ (for arbitrary $N_{\rm real}$ and $N_{\rm int}$), is trivial: The wave function of each molecule decouples from the other, and the many-body ground state is a product of single-particle ground states.

The exact ground state for (ii), $N_{\rm real}=2$ and large $N_{\rm int}$, was given in Ref.~\cite{sundar2018synthetic}, and provides much insight into the many-body physics. Summarizing, two molecules form a bound state with an even relative wave function for $V/t<0$, form a bound state with an odd relative wave function for $V/t>2$, and are in an unbound state for $0<V/t<2$. As we will see numerically in Sec.~\ref{sec: dmrg}, this picture also extends to many molecules, which form an unbound 2D gaslike state for $0\lesssim V/t \lesssim2$ and a 1D stringlike many-body bound state otherwise.

The exact solution for (iii), $t=0$, has a rich structure, and we will analytically derive it in this limit in Sec.~\ref{sec: exact}. Earlier works~\cite{iouchtchenko2018ground} have numerically studied the solution in this limit in another system of linear rotors arranged in a 1D chain. Summarizing the results of Sec.~\ref{sec: exact}, we analytically obtain the ground-state wave function and the low-energy excitations by identifying the low-energy sectors of the Hamiltonian and mapping these sectors to a spin-1/2 hardcore boson model which we solve. We find that the resulting many-body ground state spontaneously collapses to two or three synthetic sites and is therefore a tightly bound quantum string with a finite width in the synthetic dimension. The system has an intricate excitation spectrum, with gapless excitations for particle and spin fluctuations of the hardcore Bose gas, as well as gapless excitations that fluctuate the string in the synthetic direction. Remarkably, the latter excitations are gapless even though they are associated with breaking a discrete, rather than continuous, translational symmetry in the synthetic direction.

\subsection{Comparison to earlier works}
Although $\hH$ resembles an interacting 2D lattice Hamiltonian, its physics differs significantly from the physics of other lattice models such as the Hubbard model, mainly due to two important aspects: (i) The dipole interactions are off-diagonal and (ii) the allowed Hilbert space is restricted to exactly one molecule per real lattice site, which is essentially equivalent to imposing an infinitely strong and highly nonlocal interaction.

The strings in our system in fact resemble bound states or chains that were explored in some earlier works~\cite{yin2011stable, fu2012separation, capogrosso2011superfluidity, wang2006quantum}. However, our system is much more suited to prepare and observe such bound states, and the constraints required are less stringent. To create bound states of molecules, Refs.~\cite{yin2011stable, fu2012separation} consider particles trapped individually in 1D tubes or 2D sheets that extend in real space, analogous to the 1D synthetic lattice in our system. However, experimentally achieving the constraint proposed in these works -- that only one particle be present in each real space tube or sheet -- is challenging. In contrast, the same constraint arises naturally in our system by imposing a deep real space lattice and achieving unit filling in the lattice sites. The constraint in Refs.~\cite{yin2011stable, fu2012separation} can be relaxed, for example, as in Refs.~\cite{capogrosso2011superfluidity, wang2006quantum}, where they consider polarized molecular gases trapped in identical stacks, which form chains driven by dipole interactions. However, these systems will suffer from molecular collisions that lead to loss from reactions~\cite{yan2013observation, chotia2012long, zhu2014suppressing, ospelkaus2010quantum, ni2010dipolar, de2011controlling} or complicated collisional processes~\cite{mayle2012statistical, mayle2013scattering, doccaj2016ultracold, wall2017microscopic, wall2017lattice, ewart2018bosonic}. In contrast, our system avoids such processes by design, again resulting from only one molecule present per real lattice site. Moreover, the temperature requirements for Refs.~\cite{yin2011stable, fu2012separation, capogrosso2011superfluidity, wang2006quantum} are quite stringent, likely requiring much lower temperatures than have been experimentally achieved, while our system does not suffer from such stringent requirements because our molecular gas is frozen in the real space lattice, and it is straightforward to initialize the system with effectively zero entropy in the synthetic dimension. It remains an open challenge to transform easily prepared zero-entropy states into the many-body ground state, but we expect that experiments should be able to reach sufficiently cold states (with low entropy density) by an adiabatic transformation.

Our system also has significant advantages in probing the strings over the systems considered in Refs.~\cite{yin2011stable, fu2012separation, capogrosso2011superfluidity, wang2006quantum}, because straightforward spectroscopic techniques yield single-site resolution in the synthetic lattice.

\section{Matrix Product State solution for the many-body ground state}\label{sec: dmrg}
\begin{figure}[t]\centering
\begin{tabular}{cc}
(a) &\includegraphics[width=0.8\columnwidth]{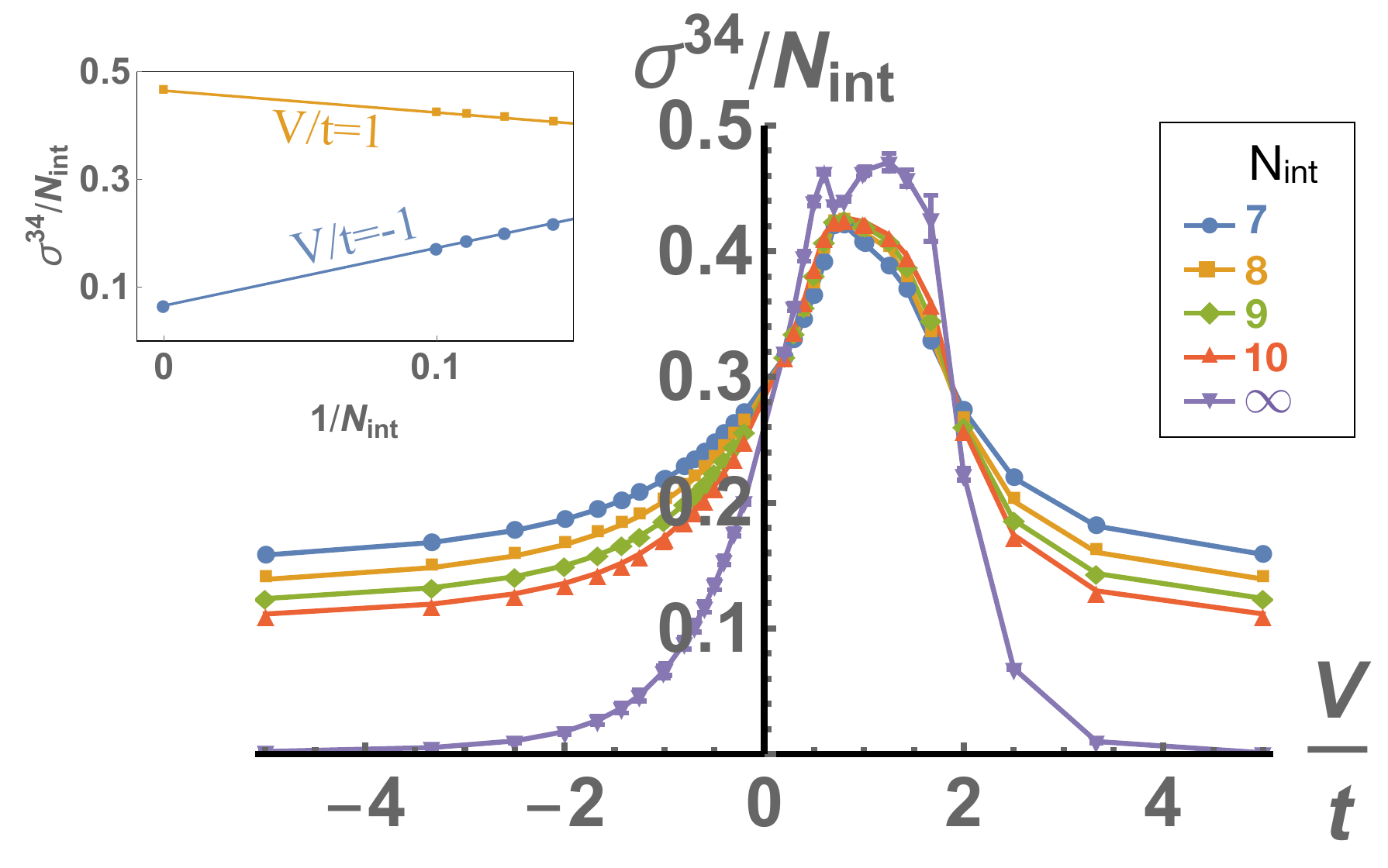}\\
(b) &\includegraphics[width=0.6\columnwidth]{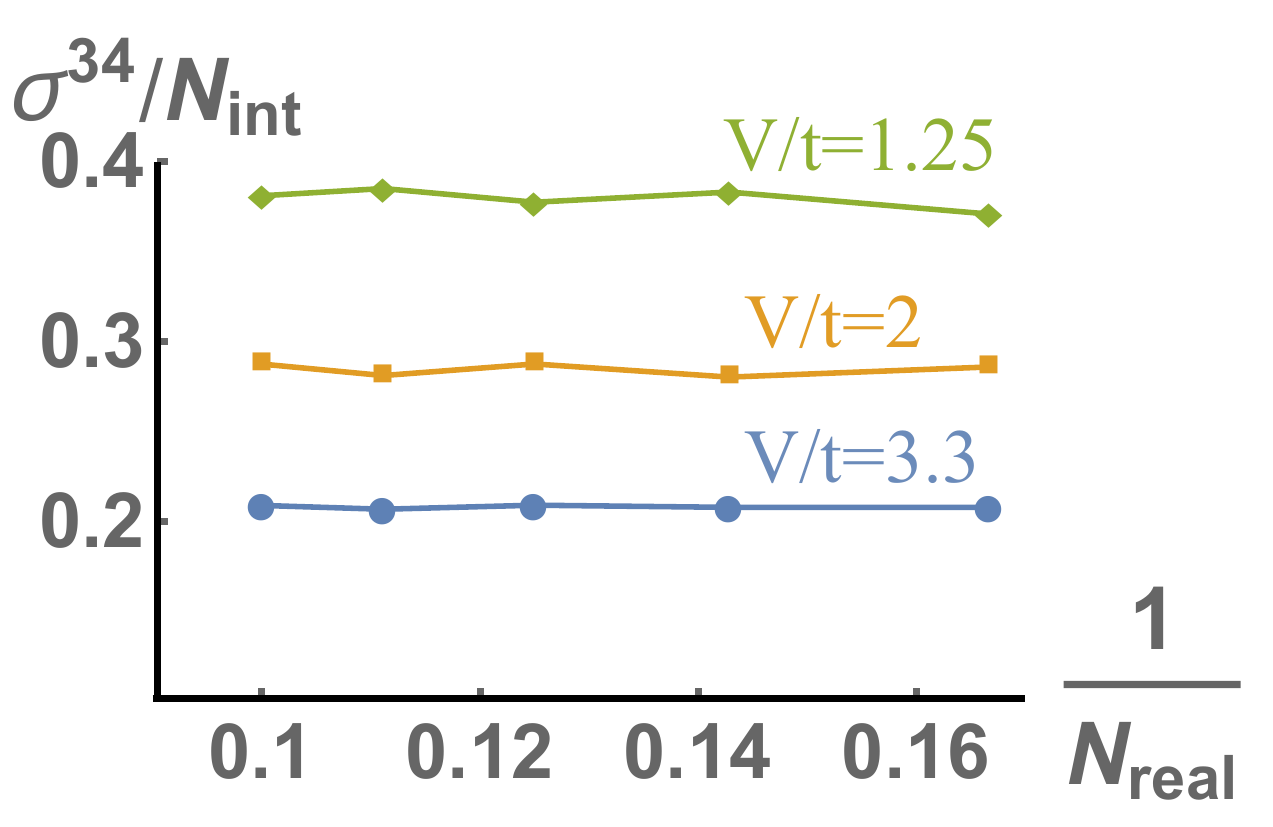}
\end{tabular}
\caption{(Color online) (a) The normalized synthetic separation, $\sigma^{34}/N_{\rm int}$, vs $V/t$ for different $N_{\rm int}$ in a 1D real-space array of $N_{\rm real}=6$ molecules. When $V/t\lesssim0$ or $V/t\gtrsim2$, $\sigma^{34}/N_{\rm int}$ is small, and the system is therefore in the string phase. The system is in an unbound phase otherwise. The curves intersect at $V/t\sim0.2$ and $V/t\sim2$, indicating that the string-gas phase transitions occur near these points. The inset shows the linear fit to the curve $\sigma^{34}/N_{\rm int}$ vs $\frac{1}{N_{\rm int}}$ used to obtain the extrapolated value of $\frac{\sigma}{N_{\rm int}}$ at $N_{\rm int}\rightarrow\infty$. The error bars indicate one standard deviation for the fit value. (b) $\sigma^{34}$ is only weakly dependent on system size $N_{\rm real}$. Blue circles, yellow squares, and green diamonds show $\sigma^{34}/N_{\rm int}$ vs $1/N_{\rm real}$ for $V/t$=3.3, 2, and 1.25 and $N_{\rm int}=6$.}
\label{fig: dmrgWidth}
\end{figure}
We calculate the many-body ground state of $\hH$ at arbitrary $V/t$ using the MPS technique~\cite{schollwock2005density, schollwock2011density, verstraete2008matrix}, implemented in the open source code openMPS~\cite{jaschke2017open}. We achieve a local energy variance $(\langle\hH^2\rangle-\langle\hH\rangle^2)/N_{\text{real}} < 10^{-14}$ in the ground state, using an adaptively chosen bond dimension whose maximum value reaches $546$. We then characterize this ground state as being extended in the synthetic dimension (i.e., in the gas phase), or having a finite width (i.e., in the string phase), by calculating the normalized separation, $\sigma^{ij}/N_{\rm int}$. Here, $\sigma^{ij}$ is the synthetic distance between two molecules at real positions $i$ and $j$:
\begin{equation}\label{eqn: sigma}
\sigma^{ij} = \sum_{m,n=0}^{N_{\rm int}-1} |m-n|\langle \hc_{mi}\+\hc_{mi}\nodag\hc_{nj}\+\hc_{nj}\nodag\rangle.
\end{equation}
In the string phases, $\sigma^{ij}$ is finite, so $\sigma^{ij}/N_{\rm int}=0$ in the thermodynamic limit, while in the unbound phase, $\sigma^{ij}\propto N_{\rm int}$, so $\sigma^{ij}/N_{\rm int}$ is finite.

Figure~\ref{fig: dmrgWidth} demonstrates the existence of two string phases and an unbound gas phase. Figure~\ref{fig: dmrgWidth}(a) plots $\sigma^{34}/N_{\text{int}}$ versus $V/t$ for $7\leq N_{\text{int}}\leq10$ in a chain with $N_{\text{real}}=6$ molecules. $\sigma^{34}/N_{\rm int}$ is large when $0\lesssim V/t\lesssim2$, indicating that the system is in a gas phase. $\sigma^{34}/N_{\rm int}$ is smaller for $V/t\gtrsim2$ and $V/t\lesssim0$, indicating a string phase. Remarkably, all the curves at different $N_{\rm int}$ intersect at $V/t\sim 0.2$ and $V/t\sim2$, clearly indicating the transition from the string to the gas phase near these points. The normalized separation approaches $O(1/N_{\rm int})$ as $V/t \rightarrow \pm \infty$.

We investigate finite-size effects along both the real and synthetic directions. To investigate finite-size effects in the synthetic dimension, we observe that $\sigma^{34}/N_{\text{int}}$ varies nearly linearly with $1/N_{\text{int}}$, and therefore extrapolate the results to $N_{\text{int}}\rightarrow \infty$ by doing a linear fit to the data for $7\leq N_{\rm int}\leq10$ [see inset in Fig.~\ref{fig: dmrgWidth}(a)]. The extrapolated results are plotted in Fig.~\ref{fig: dmrgWidth}(a) (inverted violet triangles) and show that $\sigma^{34}/N_{\rm int}$ extrapolates to nearly zero in the regions $V/t\gtrsim2$ and $V/t\lesssim-2$, indicating a string phase in these regions. Remnant nonzero $\sigma^{34}/N_{\rm int}$ for $-2\lesssim V/t \lesssim0$ is likely due to finite-size errors. For $0\lesssim V/t\lesssim 2$, the extrapolated $\sigma^{34}/N$ is finite, indicating a delocalized gaseous phase. Further, the intersection of the curves at $V/t\sim0.2$ and $V/t\sim2$ is a strong indication that the system is undergoing phase transitions here or nearby.
To investigate finite-size effects in the real dimension, we calculate $\sigma^{34}/N_{\text{int}}$ versus $1/N_{\text{real}}$ for $6\leq N_{\rm real}\leq10$ and fixed $N_{\text{int}}=6$. Figure~\ref{fig: dmrgWidth}(b) plots $\sigma^{34}/N_{\text{int}}$ versus $1/N_{\text{real}}$ for different $V/t$, showing that finite-size effects in the real space lattice are negligible.

While we are able to identify the string-gas phase transition by looking at $\sigma^{34}/N_{\rm rot}$, it would be illuminating to elucidate the phases and transitions by calculating or measuring other quantities as well. For example, finding that $\sigma^{ij}/N_{\rm rot}$ approaches zero for $|i-j|\gg1$ will provide a stronger confirmation of the presence of the string phase. While we are limited to $|i-j|=1$ in our numerics due to small system size, our finite-size extrapolations are consistent with this picture. Other correlations between the synthetic positions of molecules on real sites $i$ and $j$, or measuring the synthetic position of all the molecule in the presence of a field which biases one synthetic site of one molecule, may also be used to characterize the string phase. In addition to these methods, the string-gas phase transition may be identified by looking for singularities in the entanglement entropy~\cite{le2008entanglement, ramirez2018use, sundar2018complex, valdez2017quantifying} or the fidelity susceptibility~\cite{you2007fidelity, venuti2007quantum, schwandt2009quantum, albuquerque2010quantum}.

\section{Exact analytic solution at $t/V=0$}\label{sec: exact}
\begin{figure}[t]\centering
\includegraphics[width=0.97\columnwidth]{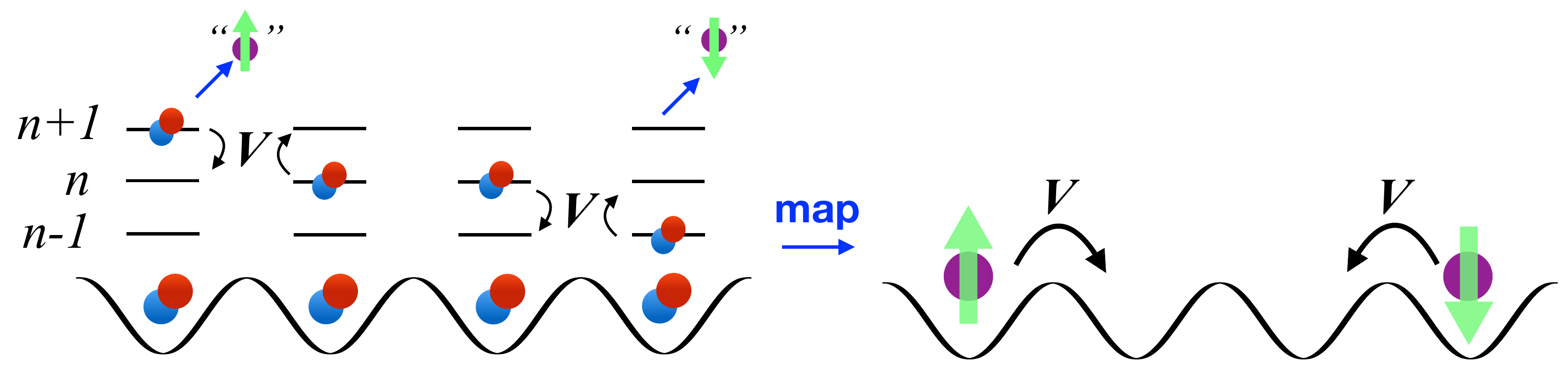}
\caption{(Color online) Schematic illustration of mapping the molecular system to hardcore bosons when $t/V=0$. $\hc_{n,j}\+\ket{\rm vac}$ maps to $\ket{\phi}_j$, and $\hc_{n+1,j}\+\ket{\rm vac}$ and $\hc_{n-1,j}\+\ket{\rm vac}$ map to $\hb_{\uparrow j}\+\ket{\phi}_j$ and $\hb_{\downarrow j}\+\ket{\phi}_j$. A dipole-induced exchange of rotational states maps to tunneling of bosons. We consider only nearest-neighbor dipole interactions, although dipole interactions are longer ranged.}
\label{fig: hardcore}
\end{figure}
In this section, we will analytically demonstrate that the ground states at $t/V=0$ are strings, and show that the strings can be either two or three synthetic sites wide. We will accomplish this by deriving an exact analytic solution for the ground-state wave function in this limit. The strings with width $3$ were absent from the mean-field theory~\cite{sundar2018synthetic}. As earlier, we work in the rotating frame, where all the synthetic sites are degenerate in energy. The ground-state solution is obtained by recognizing that $\hH$ breaks into independent sectors, and that $\hH$ maps to a hopping model for hardcore bosons in each sector that contains a ground state. In this map, the ground states correspond to condensates of the hardcore bosons. We will also shed light on the correlations and excitations in the ground state. There are $O\left(N_{\rm int}\times(2^{N_{\rm real}/2})\right)$ degenerate ground states, a signature of the richness of the system, which is also absent in the mean-field theory.

When $t/V=0$, the subspace consisting of all the molecules only on the $(n-1)$-th, $n$th and $(n+1)$-th synthetic sites is closed under the action of $\hH$, for arbitrary $n$. Clearly, any eigenstate in this subspace has a finite width in the synthetic dimension and therefore is a string. Our approach in this section will be to first to find the ground state in this subspace and then to argue that any states outside of this subspace cost energy relative to this ground state. Thus, the ground state in this subspace is the absolute ground state.

In this subspace, we map the molecular states to spin-1/2 hardcore boson states. Specifically, we map $\hc_{n,j}\+\ket{\rm vac}$ to the bosonic vacuum $\ket{\slashzero}_j$, map $\hc_{n+1,j}\+\ket{\rm vac}$ to $\ket{\uparrow}_j = \hb_{\uparrow j}\+\ket{\slashzero}_j$, and $\hc_{n-1,j}\+\ket{\rm vac}$ to $\ket{\downarrow}_j=\hb_{\downarrow j}\+\ket{\slashzero}_j$, as shown in Fig.~\ref{fig: hardcore}. The unit filling of the molecules in the real space lattice implies that two bosons cannot be on the same site $j$, and therefore these bosons can be thought to have hardcore repulsion. The Hamiltonian $\hH$ projected into the subspace with only the synthetic sites $n-1$, $n$, and $n+1$ can be rewritten in the boson Hilbert space as
\begin{equation}
\hH_n = V\sum_{j=1}^{N_{\rm real}-1}\sum_{\mu=\uparrow,\downarrow} \hb_{\mu,j}\+\hb_{\mu,j-1} + {\rm H.c.},
\label{eqn: Hprojected}
\end{equation}
with $(\hb_{\uparrow j}\+)^2 = (\hb_{\downarrow j}\+)^2 = \hb_{\uparrow j}\+\hb_{\downarrow j}\+=0$.

\subsection{Ground states}
The ground states of $\hH_n$ can be found by first mapping the spin-1/2 hardcore bosons to spinless bosons and then doing a Jordan-Wigner transformation (to review this technique, see Refs.~\cite{parkinson2010introduction, sachdev2011quantum}). All eigenstates with $M$ spin-1/2 bosons are labeled by $2M$ quantum numbers : $\mu_1\ldots\mu_M$ for their spins and $k_1\ldots k_M$ for the momenta of the Jordan-Wigner fermions. The Hamiltonian only has matrix elements between states with the same ordering of boson spins in real space, leading to each eigenstate having a fixed real-space ordering of spins, because the dipole interaction cannot swap the real space positions of adjacent molecules on the synthetic sites $n-1$ and $n+1$. In the subspace with a given ordering of spins, the matrix elements of the Hamiltonian are identical to that of a hopping model for spinless hardcore bosons,
\begin{equation}
\hat{\tilde H}_n = V\sum_{j=1}^{N_{\rm real}-1} \hat{\tilde{b}}_j\+\hat{\tilde{b}}_{j-1} + {\rm H.c.},
\label{eqn: Hspinlessbosons}
\end{equation}
which then maps to a fermionic Hamiltonian after a Jordan-Wigner transformation:
\begin{equation}\label{eqn: Hfermions}
\hat{\tilde H}_n = V\sum_{j=1}^{N_{\rm real}-1} \hat{f}_j\+\hat{f}_{j-1} + {\rm H.c.},
\end{equation}

First, we solve $\hat{\tilde{H}}_n$ for even $N_{\rm real}$. The fermionic band structure has $M=N_{\rm real}/2$ states with negative energy. Therefore, the many-body ground state of Eqs.~\eqref{eqn: Hspinlessbosons} and~\eqref{eqn: Hfermions} is the half-filled Fermi sea, given by
\begin{align}\label{eqn: gFermi}
&\ket{G} = \left(\frac{2}{N_{\rm real}+1}\right)^{M/2} \sum_{1\leq x_1< \ldots< x_M\leq N_{\rm real}} \nonumber\\
&\times A(k_1 \ldots k_M,x_1 \ldots x_M) \hat{f}_{x_1}\+ \ldots \hat{f}_{x_M}\+ \ket{\slashzero},
\end{align}
where 
\begin{equation} A(k_1 \ldots k_M,x_1 \ldots x_M) = 
\left| \begin{array}{ccc}\sin k_1x_1 & \hdots & \sin k_1x_M\\ \vdots&\ddots&\vdots\\ \sin k_Mx_1 & \hdots & \sin k_Mx_M\end{array}\right|
\end{equation}
is the Slater determinant. The momenta in Eq.~\eqref{eqn: gFermi} are
\begin{equation}
k_n = \frac{n\pi}{N_{\rm real}+1} + \frac{M\pi}{N_{\rm real}+1}\Theta(V),
\end{equation}
where $\Theta$ is the unit step function.
Rewriting Eq.~\eqref{eqn: gFermi} in terms of the spinless hardcore bosons, we obtain
\begin{align}\label{eqn: gSpinlessBose}
&\ket{G} = \left(\frac{2}{N_{\rm real}+1}\right)^{M/2} \sum_{1\leq x_1< \ldots< x_M\leq N_{\rm real}} \nonumber\\
&\times A(k_1 \ldots k_M,x_1 \ldots x_M) \hat{\tilde{b}}_{x_1}\+ \ldots \hat{\tilde{b}}_{x_M}\+ \ket{\slashzero},
\end{align}
where we have omitted the Jordan-Wigner strings attached to the bosons, as they evaluate to 1 because the sum runs over a fixed ordering of the bosons' positions $x_1< \ldots< x_M$.

Finally, the ground states for Eq.~\eqref{eqn: Hprojected} can be obtained by replacing $\hat{\tilde{b}}_{x_i}$ in Eq.~\eqref{eqn: gSpinlessBose} with $\hb_{\mu_i,x_i}$:
\begin{align}\label{eqn: gBose}
&\ket{G_{\mu_1 \ldots \mu_M}} = \left(\frac{2}{N_{\rm real}+1}\right)^{M/2} \sum_{1\leq x_1< \ldots< x_M\leq N_{\rm real}}\nonumber\\
&\times A(k_1 \ldots k_M,x_1 \ldots x_M) \hb_{\mu_1,x_1}\+ \ldots \hb_{\mu_M,x_M}\+ \ket{\slashzero}.
\end{align}

The ground states for odd $N_{\rm real}$ have the same form as Eq.~\eqref{eqn: gBose}, except that the number of bosons is $M=(N_{\rm real}\pm1)/2$ instead of $N_{\rm real}/2$.

The states in Eq.~\eqref{eqn: gBose} with all spins as $\ket{\uparrow}$ or all the spins as $\ket{\downarrow}$ are strings that are two synthetic sites wide, while all other spin combinations are strings that are three synthetic sites wide. The total number of width-$2$ ground states for even $N_{\rm real}$ is $N_{\rm int}-1$, since we can pick different values of $n$. The degeneracy for odd $N_{\rm real}$ is $2(N_{\rm int}-1)$, since the ground states can have $(N_{\rm real}-1)/2$ or $(N_{\rm real}+1)/2$ bosons. The total number of width-$3$ ground states is $(N_{\rm int}-2)(2^{N_{\rm real}/2}-2)$ for even $N_{\rm real}$, where the exponential factor is due to the different combinations of spins in real space. The number of width-$3$ ground states is $(N_{\rm int}-2)(2^{(N_{\rm real}-1)/2}+2^{(N_{\rm real}+1)/2}-4)$ for odd $N_{\rm real}$.

The only ground states of $\hH$ at $t/V=0$ are those given in Eq.~\eqref{eqn: gBose}, which are all width-$2$ or width-$3$ strings. All states occupying four or more distinct states in the synthetic dimension have a strictly higher energy than these ground states. A rigorous proof of this statement is given in Appendix \ref{appendix:A}. Summarizing the proof, suppose an eigenstate contains a molecule on the synthetic site $n+2$ in a background of molecules on synthetic sites $n-1$, $n$ and $n+1$. Note that the molecule on $n+2$ cannot resonantly swap synthetic positions with other molecules on $n-1$. Construct a new eigenstate with all molecules on $n+2$ replaced by molecules on the site $n$; in this new eigenstate, which has width 3, the earlier disallowed swap process is now allowed. Therefore, this new eigenstate (with width 3) must have a lower energy (than the original state with width 4), and thus a width-4 state cannot be the ground state. As a result, the only many-body ground states at $t/V=0$ are strings with synthetic width 2 or 3.

\subsection{Ground-state correlations}
Next, we calculate $\sigma^{ij}$ for the ground states. We do this by first writing $\sigma^{ij}$ [Eq.~\eqref{eqn: sigma}] in terms of the hardcore bosons. The only terms in the wave function [Eq.~\eqref{eqn: gBose}] that contribute to $\sigma^{ij}$ have a $\ket{\uparrow}$ boson at $i$ and $\ket{\slashzero}$ at $j$, or $\ket{\downarrow}$ at $i$ and $\ket{\slashzero}$ at $j$, or $\ket{\uparrow}$ at $i$ and $\ket{\downarrow}$ at $j$, or vice versa. Formally, this can be written as
\begin{align}\label{eqn: sigmaString}
\sigma^{ij}_{\mu_1\ldots\mu_M} =& \bra{ G_{\mu_1\ldots\mu_M} } \hb_{\uparrow i}\+\hb_{\uparrow i}\nodag (1-\hb_{\uparrow j}\+\hb_{\uparrow j}\nodag-\hb_{\downarrow j}\+\hb_{\downarrow j}\nodag)\nonumber\\
&+ \hb_{\downarrow i}\+\hb_{\downarrow i}\nodag (1-\hb_{\uparrow j}\+\hb_{\uparrow j}\nodag-\hb_{\downarrow j}\+\hb_{\downarrow j}\nodag) + 2\hb_{\uparrow i}\+\hb_{\uparrow i}\nodag \hb_{\downarrow j}\+\hb_{\downarrow j}\nodag \nonumber\\
&+ (i \leftrightarrow j)
\ket{ G_{\mu_1\ldots\mu_M} }.
\end{align}
After some cancellations, and using the fact that the real-space density $\langle \sum_{\mu=\uparrow,\downarrow} \hb_{\mu i}\+\hb_{\mu i}\nodag \rangle = \frac{1}{2} +O(1/N_{\rm real})$, Eq.~\eqref{eqn: sigmaString} simplifies to
\begin{equation}
\sigma^{ij}_{\mu_1\ldots\mu_M} = 1 - 2\sum_{\mu=\uparrow,\downarrow}\bra{ G_{\mu_1\ldots\mu_M} } \hb_{\mu i}\+\hb_{\mu i}\hb_{\mu j}\+\hb_{\mu j} \ket{ G_{\mu_1\ldots\mu_M} }.
\end{equation}
Any further simplification requires us to know the boson spins in the ground state.

For the simplest case, which is a width-$2$ string with all spins as $\ket{\uparrow}$ (or all spins as $\ket{\downarrow}$),
\begin{align}
\sigma^{ij}_{\uparrow\ldots\uparrow} =& 1 - 2\bra{ G_{\uparrow\ldots\uparrow} } \hb_{\uparrow i}\+\hb_{\uparrow i}\hb_{\uparrow j}\+\hb_{\uparrow j} \ket{ G_{\uparrow\ldots\uparrow} }\nonumber\\
=& 1 - 2\bra{G} \hat{f}_i\+\hat{f}_i\hat{f}_j\+\hat{f}_j \ket{G}.
\end{align}
The latter expression can be evaluated using Wick's theorem when $N_{\rm real}$ is large, and yields
\begin{equation}\label{eqn: sigma for width2}
\sigma^{ij}_{\uparrow\ldots\uparrow} = \frac{1}{2}+\frac{2\sin^2[\frac{\pi}{2}(i-j)]}{\pi^2(i-j)^2}.
\end{equation}

For an arbitrary width-$3$ string, we can obtain bounds for $\sigma^{ij}$ when $i$ and $j$ are sufficiently separated and $N_{\rm real}$ is large. In this limit, the densities of the bosons at $i$ and $j$ are independent and translationally invariant, resulting in $\sigma^{ij} \simeq 1-2\sum_\mu \langle \hb_\mu\+\hb_\mu\nodag\rangle^2$. Using the relation $\sum_\mu \langle \hb_\mu\+\hb_\mu\nodag\rangle \approx 1/2$, we can rewrite $\sigma^{ij}$ for large $i-j$ as
\begin{align}
\lim_{i-j\rightarrow\infty} \sigma^{ij} = &1-2\left(\langle\hb_\uparrow\+\hb_\uparrow\nodag\rangle^2 + (\frac{1}{2}-\langle\hb_\uparrow\+\hb_\uparrow\nodag\rangle)^2\rangle \right) \nonumber\\
=& \frac{3}{4} - 4 \left( \langle\hb_\uparrow\+\hb_\uparrow\nodag\rangle-\frac{1}{4}\rangle \right)^2.
\end{align}
Further, since $0\leq \langle\hb_\uparrow\+\hb_\uparrow\nodag\rangle \leq 1/2$, we can bound $\sigma^{ij}$ as
\begin{equation}
\frac{1}{2}< \lim_{i-j\rightarrow\infty} \sigma^{ij}<\frac{3}{4}.
\end{equation}

 \subsection{Excitations}
The ground states at $t/V=0$ have several striking features: They have a finite width (of 2 or 3), spontaneously collapse in the synthetic direction, and have an exponential degeneracy. As a result, the system has an intricate excitation spectrum, with at least three kinds of gapless excitations.

The first kind of excitations arise because the ground states break the discrete translational symmetry in the synthetic direction by spontaneously picking a value of $n$. As a result, the system with the string localized at the synthetic sites $n-1$, $n$, and $n+1$ has excitations that have molecules on synthetic sites $n'>n+1$ or $n'<n-1$. Remarkably, these excitations are gapless (as shown in Appendix \ref{appendix:B}), even though the symmetry broken here is discrete rather than continuous. These excitations give rise to tension in the string.

The second and third kinds of excitations are gapless particle and hole excitations associated with adding a hardcore boson to or removing a boson from the ground state. In the language of Jordan-Wigner fermions, these excitations add a fermion to or remove one from the Fermi surface, and they are gapless because the Fermi energy is zero. As an example, the excited state with one additional boson is given by
\begin{align}
&\ket{\psi} = \left(\frac{2}{N_{\rm real}+1}\right)^{(M+1)/2} \sum_{1\leq x_1< \ldots< x_{M+1}\leq N_{\rm real}}\nonumber\\ &\times A(k_1 \ldots k_{M+1},x_1 \ldots x_{M+1}) \hb_{\mu_1,x_1}\+ \ldots \hb_{\mu_{M+1},x_{M+1}}\+ \ket{\slashzero},
\end{align}
where all but one momentum are less than (greater than) $\pi/2$ for $V<0$ ($V>0$). The additional boson has a momentum greater than (less than) $\pi/2$ for $V<0$ ($V>0$), and could have either spin $\mu=\uparrow$ or $\downarrow$.

The presence of gapless excitations suggests that the string phase may be sensitive to perturbations. Furthermore, the gapless excitations are associated with a discrete symmetry breaking (rather than a continuous symmetry breaking), making their appearance somewhat mysterious. One possibility is that the model is tuned to a critical point. However, we show in the next section that the system does not seem to be critical, by demonstrating that the strings and the gas phase are robust when the interactions are modified.

\section{Dependence of the phase diagram on the form of interactions}\label{sec: electric}
\begin{figure}[t]\centering
\begin{tabular}{ll}
(a) &\includegraphics[width=0.8\columnwidth]{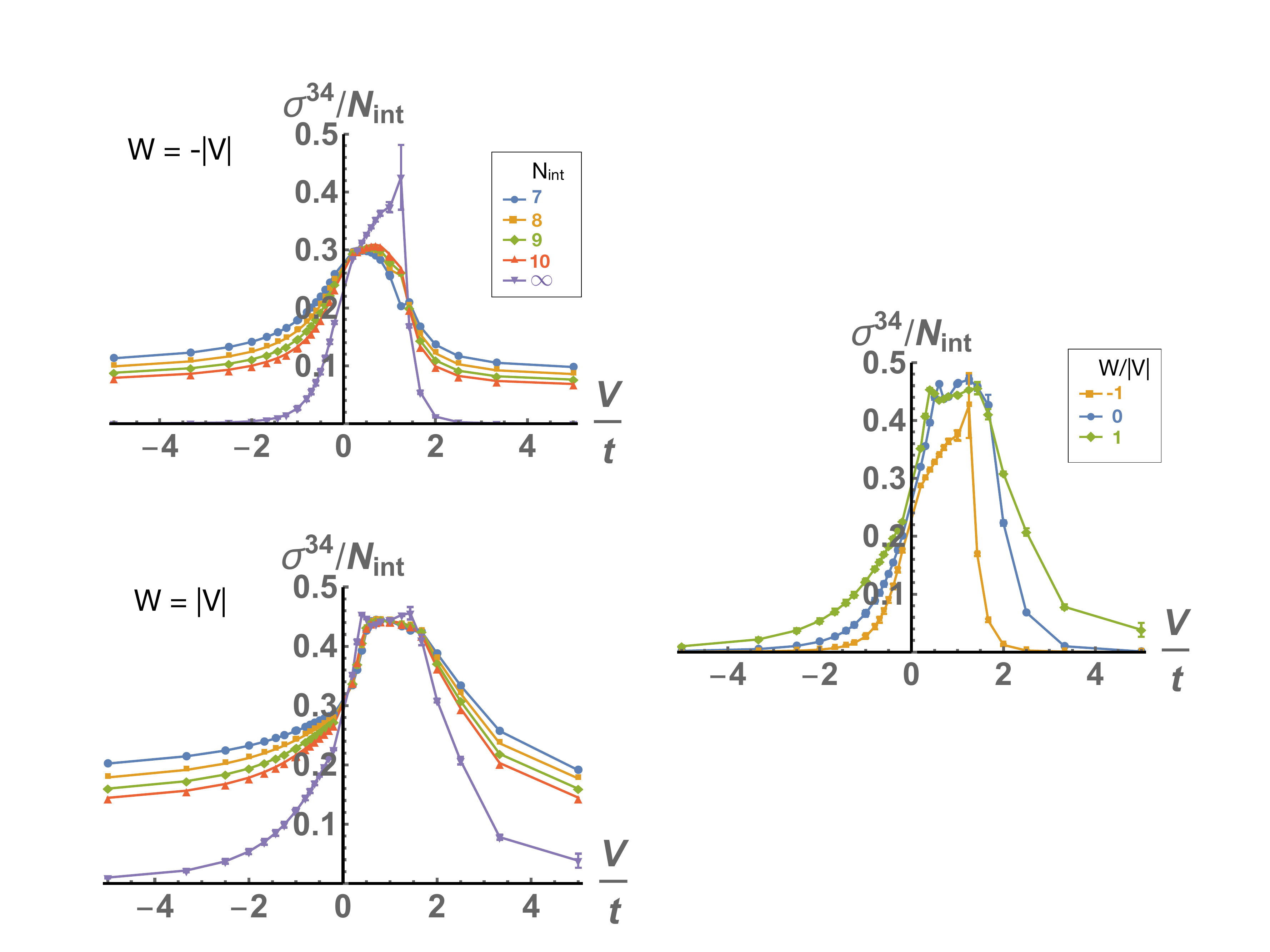}\\
(b) &\includegraphics[width=0.8\columnwidth]{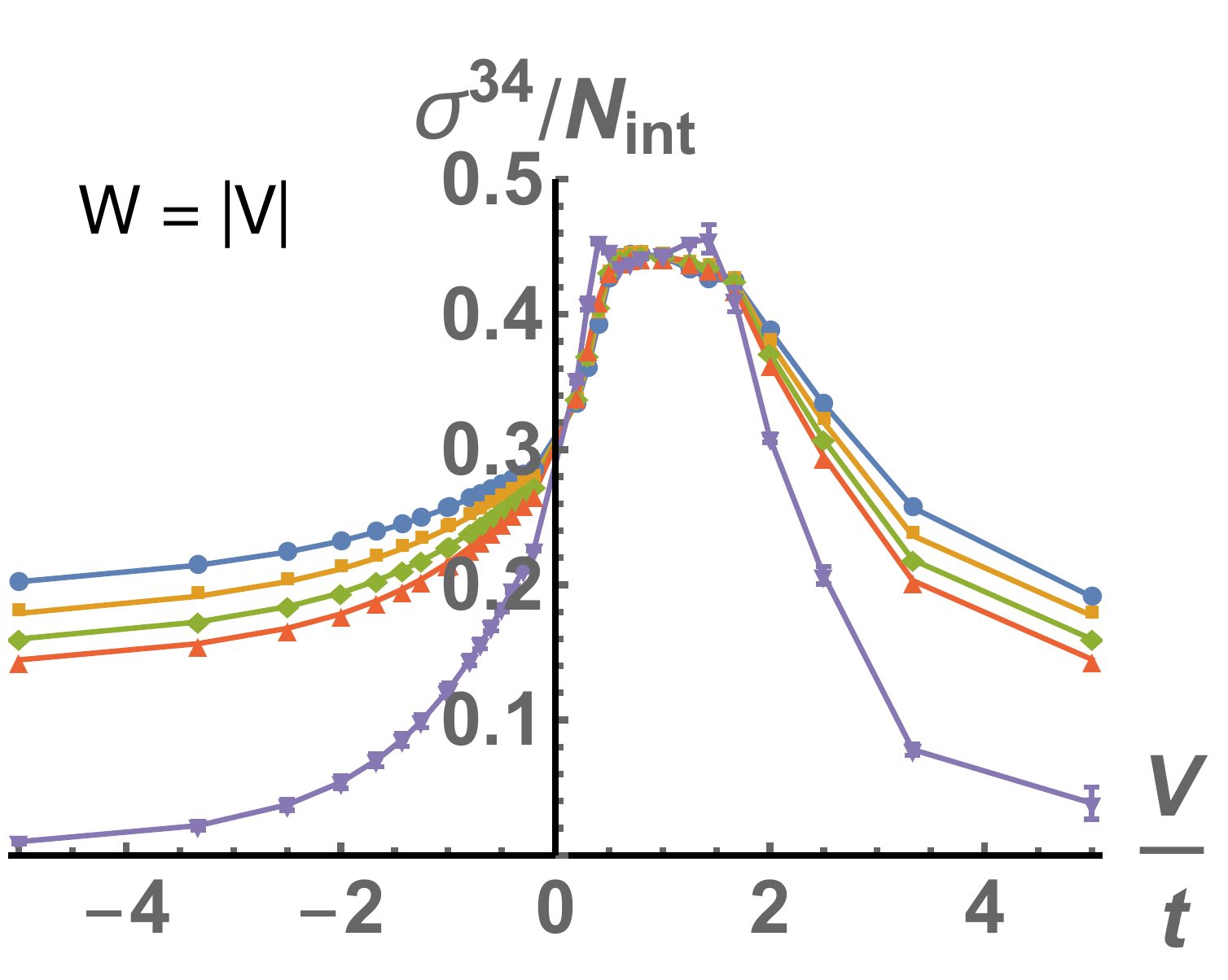}\\
(c) &\includegraphics[width=0.8\columnwidth]{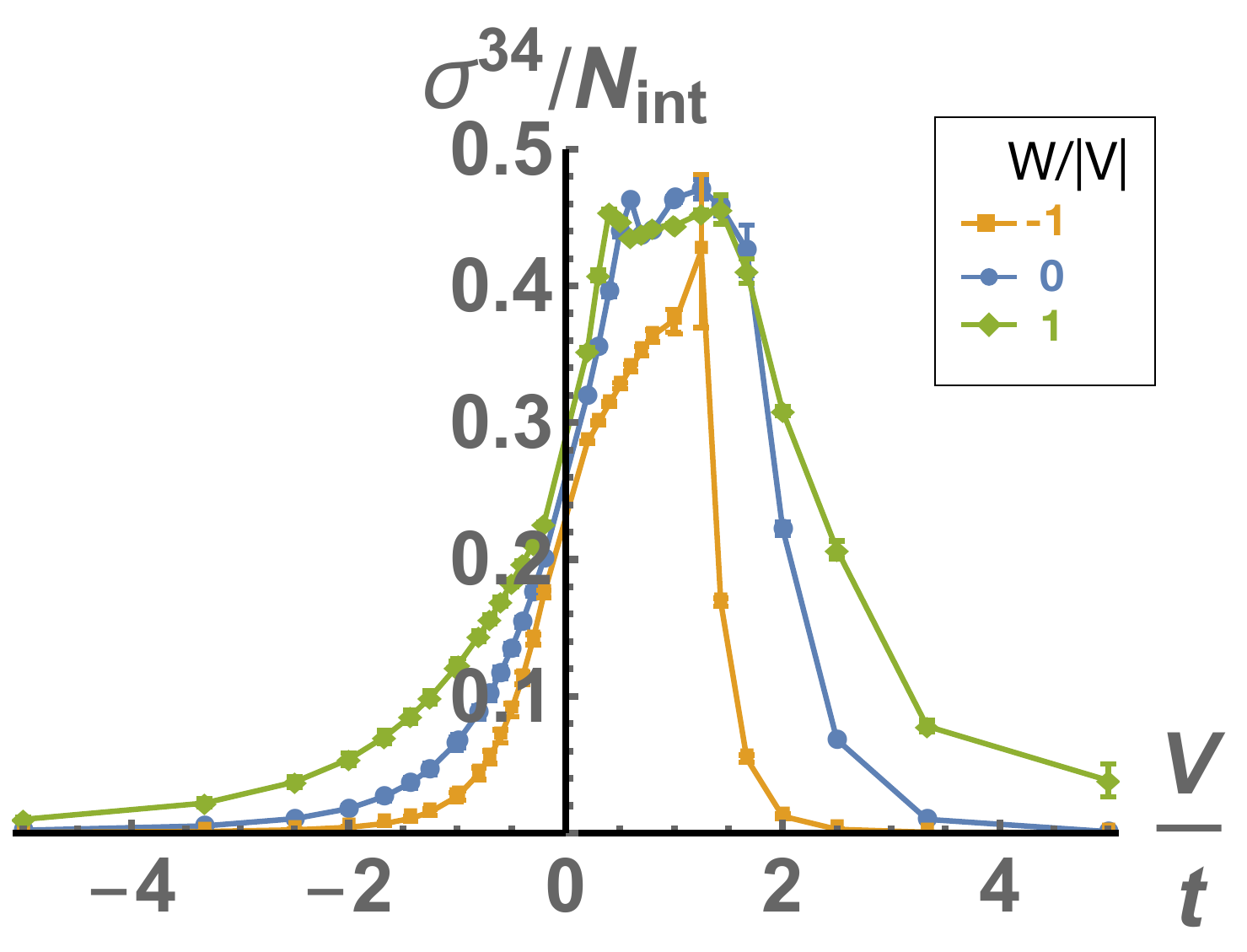}
\end{tabular}
\caption{(Color online) The normalized synthetic separation, $\sigma^{34}/N_{\rm int}$ in a 1D real-space array of $N_{\rm real}=6$ molecules and different values of $N_{\rm int}$, when (a) $W=-|V|$, and (b) $W=|V|$. The extrapolated values at $N_{\rm int}\rightarrow\infty$ are obtained by doing a linear fit to $\sigma^{34}/N_{\rm int}$ vs $1/N_{\rm int}$ as in Fig.~\ref{fig: dmrgWidth}, and the error bars indicate one standard deviation for the fit values. (c) The separation increases monotonically with $W$. Yellow squares, $W=-|V|$; blue circles, $W=0$; and green diamonds, $W=|V|$.}
\label{fig: dmrgWidth2}
\end{figure}

We modify the interactions to a form that we expect will gap some of the gapless excitations, and thus should ensure the system is in a stable phase.
Specifically, we numerically calculate the phase diagram for the Hamiltonian
\begin{equation}\label{eqn: Helec}
\hH' = \hat{H} + \sum_{nj} W \hc_{nj}\+\hc_{nj}\nodag \hc_{n,j+1}\+\hc_{n,j+1}\nodag.
\end{equation}
We find that  that the properties found previously are changed little by this modification, showing that the string ground states constitute a stable phase. The additional interactions in Eq.~\eqref{eqn: Helec} typically arise when the rotational states are dressed by external fields, and may already be present in many experiments as a result of ambient electric or magnetic fields, although generically, such fields will give rise to some additional terms as well as result in $n$-dependent $W$.

For $W<0$, we expect the molecules to be bound tighter in the synthetic direction, and the ground-state strings to only have width 2. This is clear at least for $t/V=0$ and perturbatively small negative $W$. In this limit, width-2 strings lower their energy more than any strings that are wider. Therefore, at $t/V=0$, the ground states have only width 2, and the ground-state degeneracy is $N_{\rm int}-1$. The exponential degeneracy is lifted, and the system will be stable to perturbations. We expect a similar tightening of the strings (i.e., narrower ground states) for $t/V\neq0$ and stronger attractive $W$ as well. Analogously, for $W>0$, we expect the ground states to be wider.

We demonstrate the above perturbative argument to be generally true using a numerical solution obtained from the MPS method. Figure~\ref{fig: dmrgWidth2}(a) plots $\sigma^{34}/N_{\rm int}$ versus $V/t$ at $W=-|V|$ for an array of $N_{\rm real}=6$ molecules and several values of $N_{\rm int}$. Figure~\ref{fig: dmrgWidth2}(b) plots the same quantity at $W=|V|$. Similar to Fig.~\ref{fig: dmrgWidth}, the system is in the string phase when $V/t\lesssim0$ and $V/t\gtrsim2$, and the unbound phase otherwise. At $t/V=0$, we verified numerically that the only ground states for $W<0$ have synthetic width 2, and the energy gap to width-3 strings increases proportional to $|W|$.

Figure~\ref{fig: dmrgWidth2}(c) demonstrates our argument that the string width monotonically increases with $W$, by plotting $\sigma^{34}/N_{\rm int}$ at the extrapolated $N_{\rm int}\rightarrow\infty$ for three different values of $W$. The string-gas phase transitions get sharper as $W$ decreases, and the gas phase appears in a narrower window. The sharp string-gas transition at $V/t\sim2, W=-|V|$ indicates that it is possibly a first-order transition, while the string-gas transition at $V/t\sim0$ might be of second order, although these observations must be interpreted cautiously as we are dealing with only $N_{\rm real}=6$ molecules.

\section{Summary}\label{sec: summary}
We calculated the many-body ground state of a real space one-dimensional array of polar molecules illuminated by several microwaves that drive synthetic tunnelings between sites on a lattice created by rotational states. We found numerical and analytical evidence of an intriguing phase of matter where all the molecules bind together in the synthetic direction, forming a string with rich properties, consistent with earlier mean-field predictions. However, in addition to the mean-field predictions, we showed that in the limit of strong dipole interactions, the system has exponentially many degenerate ground states with synthetic width 2 or 3, each of which can be mapped to a condensate of spin-1/2 hardcore bosons on a real-space lattice. We gave analytical expressions for these ground states and shed light on ground-state correlations and the excitation spectrum. The excitation spectrum is rich, with a number of physically distinct gapless excitations -- fluctuations of the spin-1/2 hardcore gas living on the string and excitations to states that are four synthetic sites wide. We showed that the strings constitute a robust phase and seem to be further stabilized by modified interactions.

The confluence of the numerical evidence, analytical solution at a point, and understanding of the phase in the presence of perturbations, as well as consistency of the phase diagram in our numerical results and earlier mean-field theory, all point strongly to the presence of a string phase. Besides the numerical and analytical evidence presented here, we have also discussed other methods to confirm the presence of the string phase as well as identify the string-gas phase transition, such as finding singularities in the entanglement entropy~\cite{le2008entanglement, ramirez2018use, sundar2018complex, valdez2017quantifying} or the fidelity susceptibility~\cite{you2007fidelity, venuti2007quantum, schwandt2009quantum, albuquerque2010quantum}.

Our exact numerical and analytic solutions open avenues to more thoroughly understand polar molecules with a synthetic dimension than earlier mean-field treatments, especially the nature of the string-gas phase transition, and the low-energy effective theory that describes the strings. Our analytic solutions hint at an unusual low-energy theory in which gapless excitations arise from breaking a discrete symmetry rather than a continuous one. A thorough exploration of the low-energy theory will shed light on the stability of the string phase and the fate of the gapless excitations in the presence of dipole interactions with finite strength and long range in real space.

All the physics studied in this work is immediately accessible in experiments on ultracold polar molecules, and all the physics emerges naturally with no fine-tuning. The energy scales, set by the dipole interactions, are large, and the experimental lifetimes are long due to the absence of double occupancies in the real space lattice. It will be fascinating to perform experiments to observe the quantum strings and answer the questions still left open by this work, specifically the stability of the strings, the nature of the excitations, and the nature of the phase transitions.

\section*{Acknowledgments}
This material is based upon work supported with funds from the Welch Foundation, Grant No. C-1872. K.R.A.H. thanks the Aspen Center for Physics, supported by the National Science Foundation Grant No. PHY-1066293, for its hospitality while part of this work was performed. B.G. acknowledges support by the Office of Naval Research (ONR Award No. N00014-16-1-2895).

B.S, M.T and Z.W contributed equally to this work.

\appendix
\section{Proof that the ground states of width-4 strings have higher energy than width-3 strings}\label{appendix:A}
At the end of Sec.~\ref{sec: exact}, we claimed that all ground states of the Hamiltonian $\hat{H}$ are of width 2 or 3, whose analytic expressions are given in Eq.~\eqref{eqn: gBose}, and all strings of width 4 or wider have higher energies. The current appendix is devoted to present a rigorous proof for the latter point.  

The proof involves a series of steps, mapping the original problem of a string of width 4 or greater onto a more tractable problem that will look like the width-3 problem but with certain types of bosons spatially confined. Before giving the proof, we will sketch the argument in a specific case. The proof will then generalize this argument and fill in the details. 

We obtain some insight into our proof by first taking the concrete example of a width-4 string. Let us denote the state $\prod_j \hat{c}_{n_j j}^\dagger \ket{\text{vac}}$ as $\ket{n_1 n_2 \ldots n_{N_\mathrm{real}}}$, and refer to a molecule with coordinate $(j,n_j)$ as a type-$n_j$ particle on real site $j$. Consider, for example, the width-4 string sector which includes the representative state $\ket{ \{n\} } = \ket{333222111000}$. The rest of the states in this sector can be obtained by repeatedly applying $\hat{H}$ to $\ket{ \{ n \} }$, the states that result from $\hat{H} \ket{ \{n\} }$, and so on. Our argument will first show~(Theorem~\ref{th:1}) that all the type-3 particles~(i.e., particles located on the $n=3$ site of the synthetic lattice) are restricted to a certain region of real space that is not accessible by any type-0 particle. Then we will show that it is possible to do finite steps of flipping such that after these flippings a type-3 particle is moved next to a type 0, though they can not pass through each other~(Theorem~\ref{th:2}). As a consequence, it turns out that this width-4 string maps to a width-3 string with one modification that raises its energy (Theorem~\ref{th:3}). Specifically,  particles of types 0 and 2 can both be viewed as the vacuum state, while particles of types 1 and 3 can be viewed as spin-up and spin-down hardcore bosons living on this vacuum, analogous to the solution of the width-3 string in the main text. However, the particles of type 3 now are confined in space, leading to a finite-energy cost.

In the following, we will generalize the above argument to an arbitrary sector $S$ of width-4 string states $|n_1,n_2,\ldots,n_N\rangle$ with $n_j\in\{0,1,2,3\}$, which are connected to each other by matrix elements of the interaction. We begin with a few definitions and theorems.

\begin{figure}\centering
\includegraphics[width=1\columnwidth]{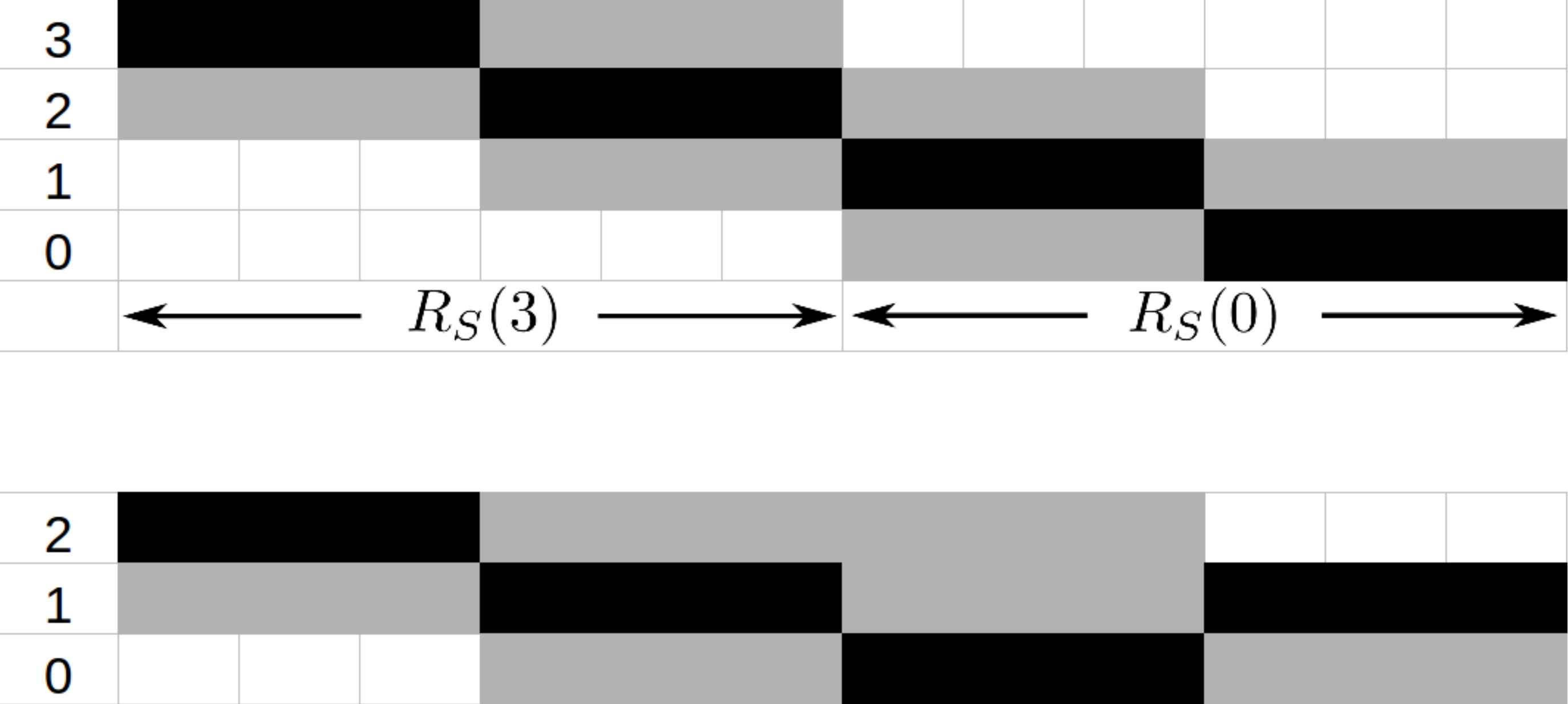}
\caption{Envelope of a width-4 string (top) compared to that of a width-3 string (bottom). The black boxes represent one configuration of molecules in the width-4 string, specifically $\ket{333222111000}$ in the top panel, and the gray boxes indicate the other types of particles that can be present in the corresponding real site in different configurations of this width-4 string sector. In this way, $R_S(n_0)$ is the intersection of the envelope with the line $n=n_0$, as shown in figure for $n_0=0,3$.}
\label{fig:RoC}
\end{figure}

\begin{figure}\centering
\includegraphics[width=1\columnwidth]{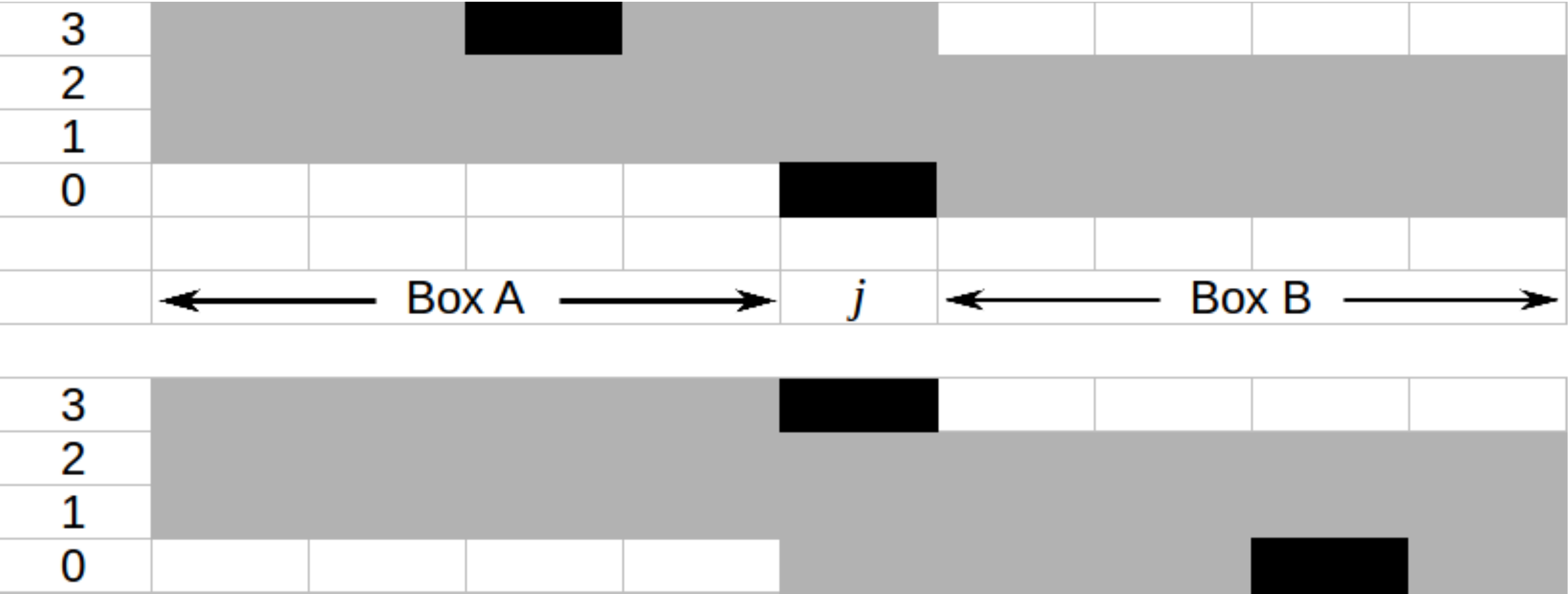}
\caption{Proof of Theorem~\ref{th:1}. Here we show two different configurations in the sector $S$, one with a type-0 particle on real site $j$~(top), the other with a type-3 particle on real site $j$~(bottom). In both cases, we denote the spatial region to the left of site $j$ as ``box A'' and the region to the right of site $j$ as ``box B.''}
\label{fig:proof1}
\end{figure}

\begin{definition} The region of confinement~(RoC) of particles of type $n$~(in the sector $S$ represented by $|n_1,n_2,\ldots,n_N\rangle$), denoted by $R_S(n)$, is defined to be the region where type-$n$ particles can possibly go after finite steps of flipping, i.e., the set of all possible positions of type-$n$ particles in this sector: $$R_S(n)=\cup_{|n_1,n_2,\ldots,n_N\rangle\in S}\{i|n_i=n\}.$$
\end{definition}
For example, for the sector with a representative state $|333222111000\rangle$, we have $R(3)=\{i|1\leq i\leq 6\}$~(since the real sites initially occupied by type 1 and type 0 can not be occupied by type 3, as type 3 can not flip with type 1 or 0) and $R(0)=\{i|7\leq i\leq 12\}$. The important observation is that the RoC of type-0 particles does not overlap with the RoC of type-3 particles. This can be generalized to the following theorem:\\
\begin{theorem}\label{th:1} $R_S(0)\cap R_S(3)=\emptyset.$
\end{theorem}
Before giving the proof, let us first comment on the indication of this theorem. For an arbitrary string state, define the envelope of this string to be the region where it can fluctuate~[e.g., the gray region in Fig.~(\ref{fig:RoC})], that is, the union of all string configurations within the same sector. The above theorem indicates that the maximal width for the envelope of an arbitrary string configuration is 3~[the width of a string envelope is defined as the maximum number of allowed particle types~(allowed synthetic states) at a real site $j$ for all $1\leq j\leq N_{\mathrm{real}}$]. Therefore, all width-3 strings trivially have a rectangular width-3 envelope, while width-4 strings have distorted envelopes. The extra energy needed to distort the envelope gives rise to an effective string tension.\\
\begin{proof} Assume the opposite, that $R_S(3)$ and $ R_S(0)$ overlap at some site $j$. Then by definition we can find two different configurations in sector $S$ with a type-0 particle and a type-3 particle occupying site $j$, respectively, as shown in Fig.~\ref{fig:proof1}. Since a type-0 particle cannot flip through a type-3 particle, the type-3 must always stand to the left of the type-0 in a given configuration. Examining Fig.~\ref{fig:proof1}, from the first configuration to the second configuration, there must be at least a particle~(e.g., of type $y$) that is removed from  box B and then added to box A. However, in order to travel from box B in the first configuration to box A in the second configuration, this type-$y$ particle must flip with both the type-0 particle and the type-3 particle~(the two black squares shown in figure). Yet no particle of any type can actually do this, a contradiction.
\end{proof}
\begin{theorem}\label{th:2} There exists at least one point $i$ where $R_S(0)$ and $R_S(3)$ touch each other, i.e., such that $i\in R_S(0),i+1\in R_S(3) $ or $i\in R_S(3),i+1\in R_S(0) $.
\end{theorem}
\begin{figure}\centering
\includegraphics[width=1\columnwidth]{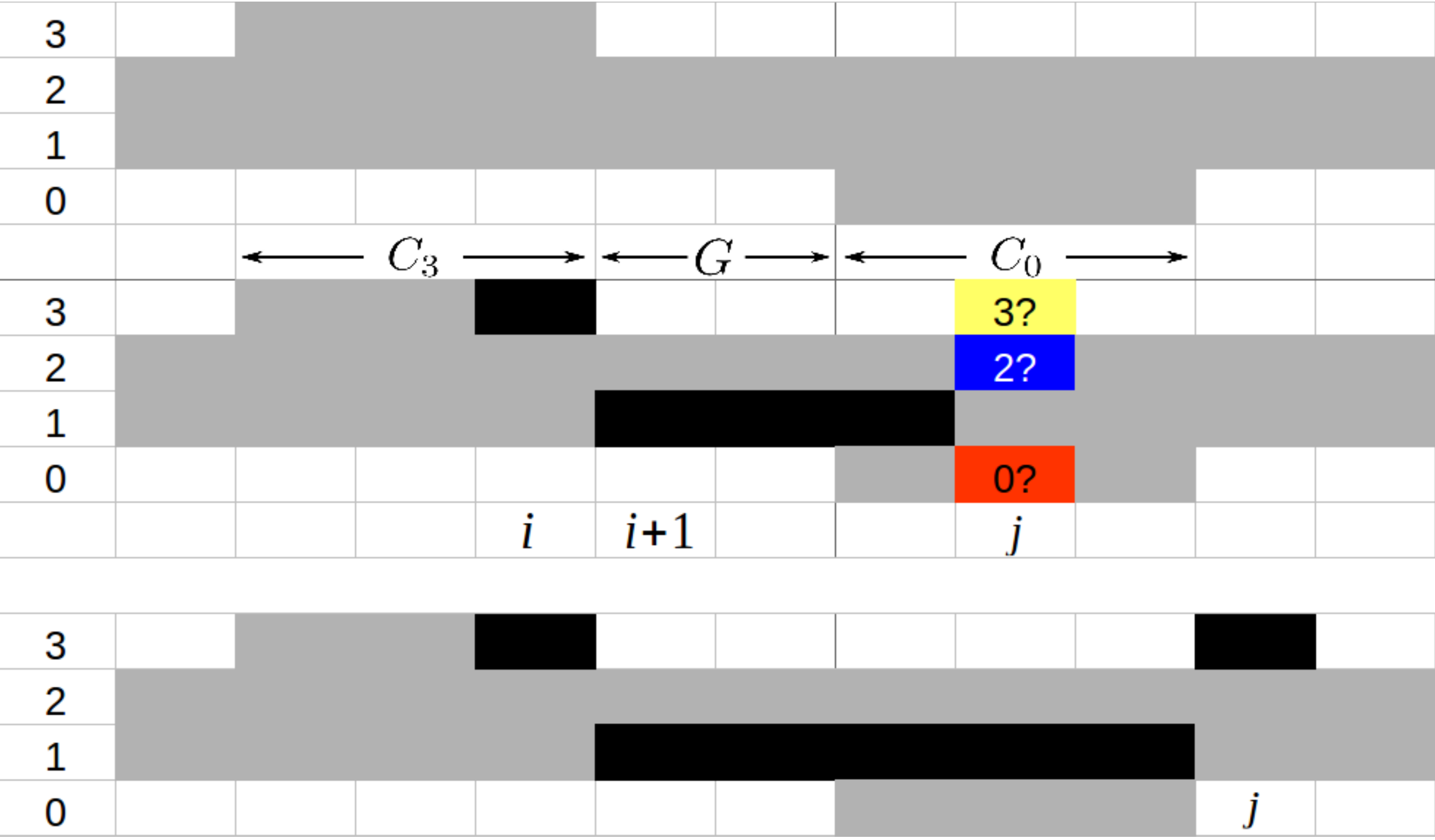}
\caption{Proof of Theorem~\ref{th:2}. Top: relative positions of $C_3,G,C_0$. Middle: $n_j$ on site $j$ is the leftmost particle on the right of site $i$ that is not a type-1. Bottom: the case $n_j=3$, and $j\notin G, j\notin C_0$.}
\label{fig:proof2}
\end{figure}
\begin{proof} Assume the opposite, that $R_S(0)$ and $R_S(3)$ do not touch anywhere. Let $C_3\subseteq R_S(3),C_0\subseteq R_S(0)$ be two maximal connected pieces of $R_S(3)$ and $R_S(0)$, respectively~[maximal means that $C_0$ is not a proper subset of any connected piece of $R_S(0)$, and similarly for $C_3$; see Fig.~\ref{fig:proof2} for an example]. Without loss of generality, suppose that $C_3$ is the rightmost connected piece of $R_S(3)$ that is on the left of $C_0$, as shown in Fig.~\ref{fig:proof2}. In this way, there must be a gap region $G\neq \emptyset$ between $C_3$ and $C_0$ where neither type-3 nor type-0 particles are allowed.

Consider a configuration where a type-3 particle occupies the site $i$, the \textit{rightmost} site of $R_S(3)$. Then the particle on site $i+1$ cannot be a type-2; otherwise, the type 3 on site $i$ can be flipped to site $i+1$. So there must be a type-1 particle on site $i+1$. Consider, then, the \textit{leftmost} particle $n_j$ on the right of site $i$ that is not a type-1, and suppose it stands on site $j$. By definition, $n_j\neq 1$. The question is what $n_j$ might be. It cannot be type-2~(blue square in Fig.~\ref{fig:proof2}); otherwise, we can flip $n_j=2$ to site $i+1$ and then flip with the type-3 on site $i$ so that the type-3 occupies site $i+1$, a contradiction.  If $n_j=0$, then we can flip $n_j$ to site $i+1$; this contradicts the fact that $i+1\notin R_S(3)$. The remaining possibility is $n_j=3$, and $j\notin G, j\notin C_3$~($j\in C_3$ would contradict Theorem~\ref{th:1}), as shown in the third configuration in Fig.~\ref{fig:proof2}. This case contradicts the fact that $C_0\subseteq R_S(0)$ in that there is no way for any type-0 to enter the region $C_0$, as type-0 from outside cannot pass through the two type-3s shown in the figure, and the region between $i$ and $j$ is filed with type-1 particles. This completes the proof.\end{proof}

Our next step is to map an arbitrary sector $S$ of width-4 string to a system of nearest-neighbor tunneling hardcore bosons where a subset of bosons are confined to a certain spatial region. We start with the Hamiltonian for the width-4 string~(acting on the sector $S$)
\begin{equation}
\hat{H}_S=\hat{E}_{01}+\hat{E}_{12}+\hat{E}_{23},
 \end{equation}
 where $\hat{E}_{01}$ is the nearest-neighbor exchange interaction between type-0 and type-1 particles and similarly for $\hat{E}_{12},\hat{E}_{23}$. Consider a modified Hamiltonian
 \begin{equation}
\hat{H}'_S=\hat{E}_{01}+\hat{E}_{12}+\hat{E}_{23}+R_S(0)+R_S(3)+\hat{E}_{03},
\end{equation}
where the term $R_S(0)$ gives infinite energy to configurations where a type-0 is outside $R_S(0)$~[i.e., the term $R_S(0)$ restricts all type-0 particles to be inside $R_S(0)$], and similarly for $R_S(3)$, and we added an exchange interaction $\hat{E}_{03}$ between type-0 and type-3. From the definition of $R_S(0)$ and $R_S(3)$, the addition of these two terms does not change the matrix elements of the Hamiltonian in the sector $S$. Furthermore, according to theorem~\ref{th:1}, $R_S(0)\cap R_S(3)=\emptyset$, in the sector $S$ where all type-0 particles are confined in $R_S(0)$ and all type -3 particles are confined in $R_S(3)$, $\hat{E}_{03}$ has zero matrix element between any two states in the sector $S$. Therefore, in the sector $S$ the new Hamiltonian $\hat{H}'_S$ has the same matrix elements with $\hat{H}_S$: $\hat{H}'_S\cong\hat{H}_S$. The string described by the Hamiltonian $\hat{H}'_S$ can be mapped to a system of two types~(1,3) of hardcore bosons tunneling in two types of vacuum~(0,2), where type-3 bosons are confined in $R_S(3)$ and type-0 vacua are confined in $R_S(0)$. Confinement on vacuum sites is not natural to deal with; therefore, our last step is to get rid of $R_S(0)$: We define
\begin{equation}\label{eq:H''}
\hat{H}''_S=\hat{E}_{01}+\hat{E}_{12}+\hat{E}_{23}+R_S(3)+\hat{E}_{03}.
\end{equation}

\begin{theorem}\label{th:3} $\hat{H}''_S\cong \hat{H}'_S$.\end{theorem}
  That is, as long as all type-3 particles are confined in $R_S(3)$, even if we are allowed to flip type-0 and type-3 by the $\hat{E}_{03}$ term, all type-0s would be \textit{automatically} confined in $R_S(0)$.

\begin{proof} Let us start with an arbitrary representative state $|S_0\rangle$ of sector $S$. Assume the opposite, that after a finite sequence of flipping with $\hat{E}_{01},\hat{E}_{12},\hat{E}_{23},\hat{E}_{03}$~[with all type-3 particles confined in $R_S(3)$], $|S_0\rangle$ is transformed to $|S_n\rangle$ where at least one type-0 particle leaves $R_S(0)$. In this sequence, $\hat{E}_{03}$ must be used at least once; otherwise, it would contradict the definition of $R_S(0)$. Consider the very \textit{first} step when a type-0 and type-3 are flipped (say, at step $r$, $|S_{r-1}\rangle\to|S_{r}\rangle$). Then, before this flipping, all type-0s must be inside $R_S(0)$~[by definition of $R_S(0)$]. Since a type-0 and a type-3 are flipped at step $r$, in the state $|S_{r-1}\rangle$ this type-0 and type-3 must be nearest neighbors, and they must be inside $R_S(0)$ and $R_S(3)$, respectively, so that the flipping occurs exactly at the touching point between $R_S(0)$ and $R_S(3)$. But in this way, this flipping would move the type-3 out of $R_S(3)$, which is forbidden in $\hat{H}''_S$, a contradiction.
\end{proof}

Without the confinement term $R_S(3)$, the Hamiltonian $\hat{H}''_S$ can be mapped to a system of two types of hardcore bosons (types 1 and 3) tunneling in a sea of two types of vacuums (types 0 and 2). Therefore, we have proved that the original Hamiltonian $\hat{H}_S$ of width-4 strings can be mapped to $\hat{H}''_S$ describing a system of nearest-neighbor tunneling hardcore bosons (of types 1 and 3) with confinement on type-3 bosons. Compared to width-3 or width-2 strings in their ground state, width-4 strings have a higher energy due to the additional confinement.

\section{The excitation energy to wider strings is exponentially small in system size}\label{appendix:B}
We finally give some arguments showing that the excitation energy from width-2 or width-3 ground state to the width-4 ground state decreases exponentially with system size $\Delta E\propto \exp(-N_{\mathrm{real}})$. For simplicity, suppose $N_{\mathrm{real}}=2M$, and consider the lowest energy state of a width-4 string sector with one type-0, one type-3, $(M-1)$ type-1, and $(M-1)$ type-2 particles, with a representative state $|011\ldots122\ldots23\rangle$.  Without the confinement term $R_S(3)$ in Eq.~\eqref{eq:H''}, the system is equivalent to a width-2 string state with $M$ hardcore bosons, so the ground-state wave function is a superposition of $\binom{2M}{M}$ terms. Going back to the width-4 string, the confinement $R_S(3)$ only forbids one component of the wave function represented by $|11\ldots 13022\ldots 2\rangle$, and the dimension of the  sector $S$ is $\binom{2M}{M}-1$. So compared to the width-2 string ground state, the modification to the width-4 string ground-state wave function and energy is proportional to $1/\binom{2M}{M}$, an exponentially small number in system size.

\bibliography{BibFile}

\end{document}